# Using Liquid Metal in an Electromechanical Motor with Breathing Mode Motion

Farhad Farzbod, Masoud Naghdi, and Paul M Goggans

*Abstract*—**Electromechanical actuators exploit the Lorentz force law to convert electrical energy into rotational or linear mechanical energy. In these electromagnetically induced motions, the electrical current flows through wires that are rigid and consequently the types of motion generated are limited. Recent advances in preparing liquid metal alloys permit wires that are flexible. Such wires have been used to fabricate various forms of flexible connections, but very little has been done to use liquid metal as an actuator. In this paper we propose and have tested a new type of motor using liquid metal conductors in which radial (or breathing) modes are activated.**

*Index Terms*—**Actuators, Breathing Mode, Galinstan, Liquid Metal, Motors.**

## I. INTRODUCTION

ELECTROMECHANICAL actuators that exploit the Lorentz force law have been used in various applications. The physics involved in these motors allows actuators to be designed having a wide range of sizes and mechanical energy output. Unlike internal combustion engines these devices produce no pollution, making them suited to applications such as indoor machinery and medical appliances. These actuators generally consist of current-carrying wires which move due to the Lorentz force. Depending on the application, the wires are made from differing metals. They are solids and not designed to be stretched or change shape. The resulting motion is purely linear or rotational.

For wearable computers, flexibility of the wires and current-carrying elements is required. Various materials have been developed to this end. Post et al. [1] developed conductive textiles for use in flexible multilayer circuits. To enhance the electrical conductivity while retaining flexibility, Lacour et al. [2] deposited a thin layer of gold having surface wrinkles on a polydimethylsiloxane (PDMS) substrate; the resulting strip was an elastic electrical conductor. Gray et al. [3] exploited the fact that metals bend and twist when their cross sections are small enough. They used twisted gold wires on PDMS substrate to produce a stretchable conductor. Siegel et al. [4] used techniques of microfluidics fabrication and injected liquid solder into microfluidic channels. When the solder cools and solidifies, the result is a flexible metallic structure. Dickey et al. [5] investigated various properties of liquid metal eutectic gallium-indium (EGaIn) injected into a microchannel. Since soft material structures, with liquid metals in their channels, are inherently less prone to mechanical fatigue and fracture, various liquid alloys have been used by other researchers. Gallium is the main element in these alloys, because it is one of only five metals that are liquid at room temperature; the others are rubidium, cesium, francium and mercury [6]. The first three are in the alkali metals group and react violently with water, while mercury is highly toxic.

Eutectic gallium-indium (EGaIn) and Galinstan are liquid metal alloys. EGaIn consists of gallium and indium whereas galinstan has a different ratio of gallium to indium, plus tin. As a result of their low toxicity [6, 7], galinstan and EGaIn have been used to fabricate various structures. Flexible interconnections [8-10] have been made using PDMS filled with galinstan. Other devices include inductors [11], a microfluidic wireless power system [12], loudspeakers and microphones [13], contact pressure sensors [14], particle-filled composite [15] and fluidic antennas [16]. In all of these applications except for the loudspeaker, the liquid metal works as a sensor or a current-carrying element of the system. In the present work the liquid metal is used as an actuator, exerting force on the surroundings. This is in essence a new kind of motor in which the 'wires' are flexible, and as such the motions they generate are not restricted to rotational or linear motion. This type of actuator utilizes the Lorentz force in the same way as conventional electromechanical motors, but we may now envision diverse configurations and geometries so as to generate very complex types of motion which are impossible to realize otherwise.

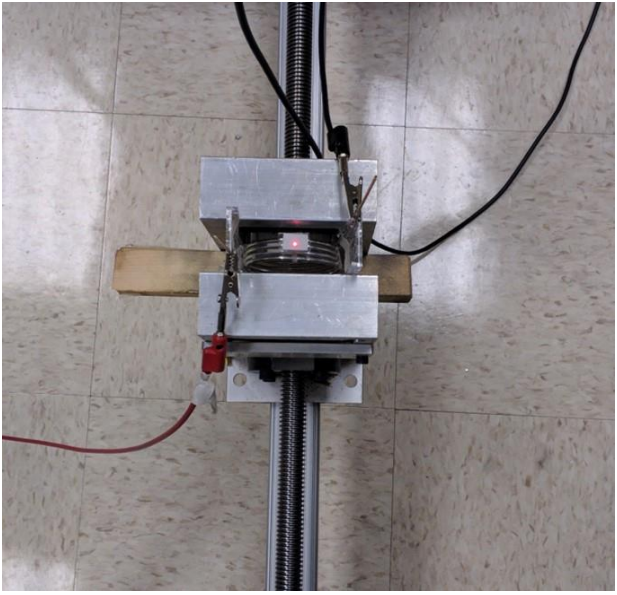

Figure 1 Helix made with PTFE vibrating in air while the out-of-plane displacement is measured with a laser Doppler vibrometer.

## II. LIQUID METAL ELECTROMECHANICAL MOTOR

### A. Principle of operation

This motor consists of a tube and two permanent magnets. The tube is made of soft elastic material in the shape of a helix, and is filled with liquid metal Galinstan (see Fig. 1). This tube is held between two permanent magnets. The current through the liquid metal is modulated using an amplifier. The generated force is given by the Lorentz force law:

$$d\mathbf{F} = id\mathbf{l} \times \mathbf{B} \quad (1)$$

in which $i$ is the current, $\mathbf{B}$ is the magnetic field and $d\mathbf{l}$ is the differential length vector parallel to the current. The differential force vectors are in the radial direction, pointing inward or outward according to the direction of the current. As a result there is no net torque on the structure. By symmetry, the net force is also zero. Since the tube is made with flexible materials, however, these forces induce vibrations in the tube, taking the form of a 'breathing' mode.

In this mechanism the mass comprises the mass of the tube and the mass of the liquid metal. The stiffness is that of the tube only. During free vibration in air, each helix turn produces the same amount of vibration force while adding the same amount of mass to the overall structure. Consequently, increasing the number of turns would not affect the displacement or the acceleration.

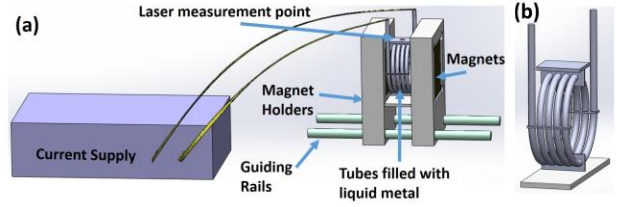

Figure 2: (a) Schematic of the liquid metal electromechanical system. (b) Helix coil constrained at the bottom. The turns are tied together in three further locations.

### B. Experimental Setup

For the tube we used three materials: polyurethane, polyvinyl chloride (PVC) and perfluoroalkoxy alkanes (PFA). All tubes had an external diameter of 1/8 in and an internal diameter of 1/16 in. The diameter of the helix was 2 inches in all cases, and there were four turns (see Fig. 2a.) The helix is fixed at the bottom by a thin structure. The helix turns are also tied together by thin hollow structures, as shown in Fig. 2b.

The magnets were off-the-shelf NdFeB Grade N52 permanent magnets, purchased from K&J Magnetics. Their dimensions were 3×3×1 in. The magnets were kept separate by an aluminum structure with a steel lead screw. The magnetic field strength was measured as 0.3 T in the middle of the space between the magnets, and is unchanged when the tubes filled with liquid metal are in place. The current was provided by a current amplifier and the input frequency range was 20 Hz to 5 kHz (500 points). As the resistance of the tube filled with Galinstan was 0.1 Ω, we use a 3.07 Ω resistor in series with the motor to reduce the load on the amplifier, and also measure the current. The peak current was set to be 0.1, 0.3, 0.5 and 0.7A, depending on the tube material. The out-of-plane displacement was measured by a Polytec PDV-100 Laser Doppler vibrometer (LDV). The liquid metal was injected into the tubes by an air pump. After the tubes were filled with liquid metal, copper tubes were pushed through the liquid to make contact with the galinstan. These copper tubes also provide space for the galinstan to move in and out of the tube when the tubes flex. To check the connection and verify the absence of disconnecting bubbles in the tube, the resistance of the tube is measured before and after the experiment.

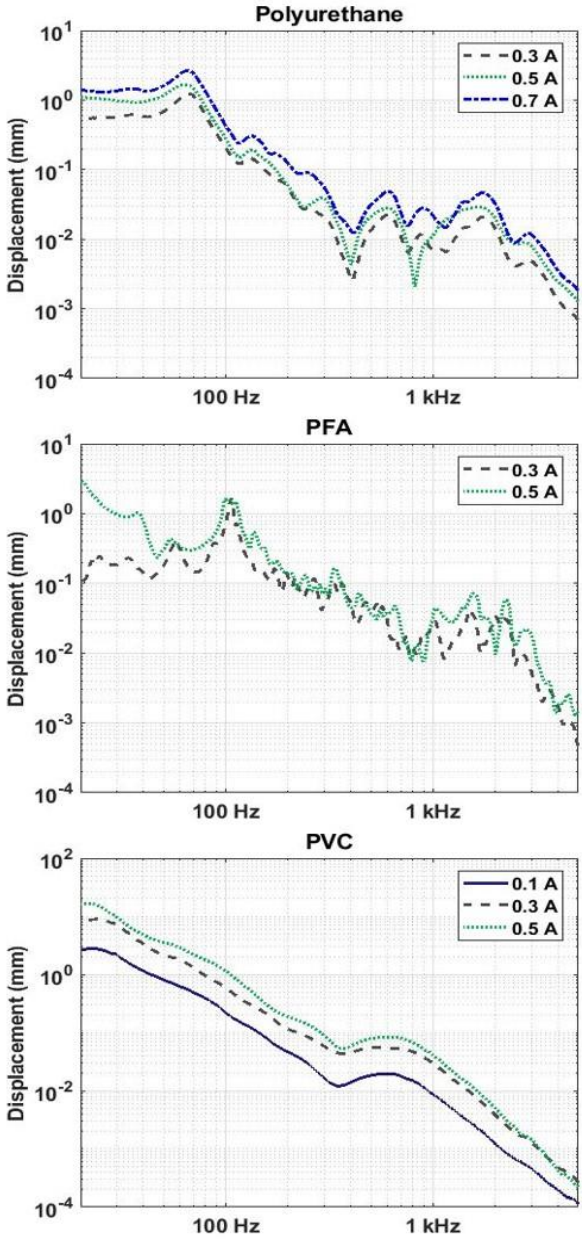

Figure 3: Displacement vs frequency for three different tube materials; Polyurethane, PFA and PVC.

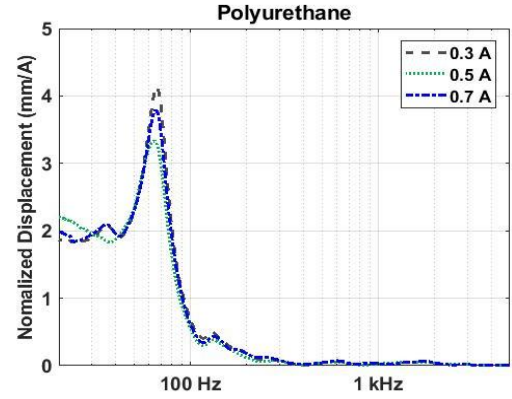

Figure 4: Normalized displacement vs frequency for the polyurethane tube.

### C. Experimental Results and Discussion

Fig. 3 shows the measured displacement plotted against frequency for the three different tube materials and various input currents. In Fig. 4, the normalized displacement is plotted on a linear scale against frequency for the polyurethane tube. The displacement varies linearly with the input current. The first resonant frequency is the strongest one in all cases. This is expected, as the distributed forces on the tubes are uniform and have radial symmetry. The first resonant frequency and its amplitude depend on the hardness of the tube (see Fig. 5). The elastic modulus of polyurethane, PFA and PVC were calculated from the hardness of the material as quoted by the manufacturer. We used the empirical relation due to Gent [17]:

$$E = \frac{0.0981(56+7.62336S)}{0.137505(254-2.54S)} \quad (2)$$

where E is the Young's modulus in MPa, and **S** denotes the type A durometer hardness. Values of the Young's modulus were calculated to be 43.83, 4.43 and 67.17 MPa for polyurethane, PVC and PFA, respectively.

### D. Modeling Results

Resonance frequencies are a highly reliable characteristic of a mechanical system. They are essentially eigenvalues for which the mechanical energy Lagrangian takes its extremum values [18]. We therefore use the experimental resonance frequencies as a criterion to validate our model. As mentioned above, as a result of the symmetry in force distribution and the geometry involved, the first resonant frequencies were excited to a greater extent in this experiment.

To model the vibration of this motor, we considered a single loop of the helix. In air, the vibration of a single loop of a helix should not differ from the vibration of multiple turns. The liquid metal moves in the tube as the Lorentz force causes the tube to vibrate. This motion of the liquid metal induces friction with the tube and there are other complex interactions caused by the fluid flow. For simplicity we neglect such effects due to movement of the liquid metal in the tube. In our model, we assume a hollow tube with the elastic constants defined in sec II-C. In calculating the mass of the tube we take into account the mass of the liquid metal as if it were distributed uniformly on the tube wall. Based on the volume of the liquid metal in one loop and its density, we calculate the mass of the liquid metal. This mass is then assumed to be distributed on the tube wall. We accordingly define a density for this distributed mass as the mass of the liquid metal divided by the volume of the wall. The density of the hollow tube in our model is then calculated as the density of the tube material plus the calculated density of the distributed material.

Table 1 sets out the experimental resonance frequency and the ones simulated from the model in COMSOL. There is excellent agreement between these frequencies for polyurethane and PVC. For PFA, however, there is approximately 25% error. We attribute this error to the way in which these tubes were connected together and to the structure, which were not totally isolated. In our experiments the coupling to the ground was stronger for the PFA because of the tightness of the fit to the pipe holders. We nevertheless see the resonant frequencies increase as the elastic constant increases. The first resonant frequency is a radial breathing mode, as expected. Figure 6 shows the first displacement eigenmode projected in the *y* direction for polyurethane.

Table 1: First resonant frequency, calculated by model and measured experimentally

|  | Polyurethane | PFA | PVC |
|---|---|---|---|
| Experimental | 67 Hz | 106 Hz | 24 Hz |
| Model | 69 Hz | 75.23 Hz | 25.19 Hz |

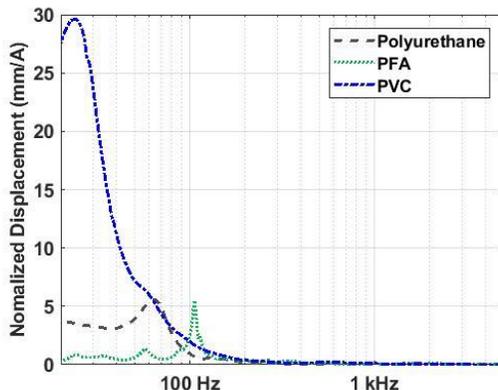

Figure 5: Normalized displacement vs frequency for the three tube materials.

III. CONCLUSION

We have investigated new types of actuators in which a breathing mode is activated. Tubes filled with liquid metal galinstan were used to study this type of motion.

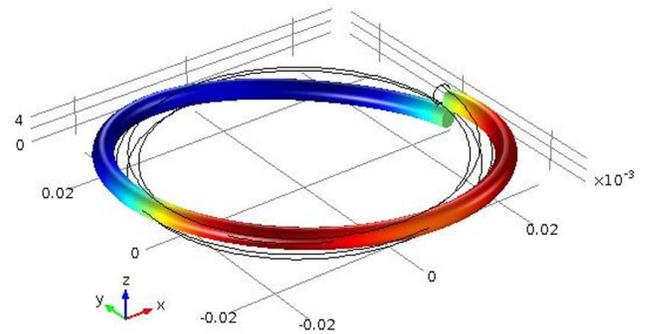

Figure 6: Heat map of the displacement eigenmode in the *y* direction associated with the first mode for polyurethane.

We used various tube materials and measured the output displacement versus frequency in the range 20 Hz to 5 kHz. The actuator has a linear response in relation to the input current. In the model, it was found that the effect of liquid metal motion can be neglected in simple designs; it is possible to make narrow tubes so as to increase the resonant frequencies by reducing the liquid metal mass and increasing the stiffness. Although we used a simple helix configuration for the tubes, and the design is simple, this type of actuator can be deployed in much more complex geometries, to study complex motions and motors. Movement of liquid mass can, for instance, be incorporated into the design by adding a reservoir for liquid metal in the tube, or by making channels of complex shape so as to focus energy within some region.